\documentstyle[12pt,psfig]{article}
\begin{document}
\pagestyle{plain}
\setcounter{page}{1}
\baselineskip16pt

\begin{titlepage}

\begin{flushright}
PUPT-1586\\
hep-th/9601057
\end{flushright}
\vspace{20 mm}

\begin{center}
{\huge Gravitational lensing by $p$-branes}

\vspace{5mm}

\end{center}

\vspace{10 mm}

\begin{center}
{\large S.S.~Gubser, A.~Hashimoto, I.R.~Klebanov, and J.M.~Maldacena}

\vspace{3mm}

Joseph Henry Laboratories\\
Princeton University\\
Princeton, New Jersey 08544

\end{center}

\vspace{2cm}

\begin{center}
{\large Abstract}
\end{center}

\noindent
The scattering of R-R gauge bosons off of Dirichlet $p$-branes is 
computed to leading order in the string coupling.  
The results are qualitatively similar to those found in the scattering
of massless NS-NS bosons: all $p$-branes with $p\geq 0$ exhibit
stringy properties, in particular the Regge behavior.
Both the R-R and NS-NS scattering amplitudes agree in the limit of small
momentum transfer with scattering off the extremal
R-R charged $p$-brane solutions found in the low-energy
supergravities.  We interpret this as
evidence that Dirichlet-branes are an exact world-sheet description of
the extremal $p$-branes. The $-1$-brane (D-instanton) is a special
object which, unlike all other Dirichlet-branes, exhibits point-like
behavior. We show that
the field theoretic scattering off of 
the R-R charged instanton solution to type~IIB supergravity
exactly reproduces the full stringy calculation.
As an aside, we include a discussion of the entropy of non-extremal
black holes in ten dimensions, produced by exciting the $0$-brane.

\vspace{2cm}
\begin{flushleft}
January 1996
\end{flushleft}
\end{titlepage}
\newpage
\renewcommand{\baselinestretch}{1.1}  



\newcommand{\fslash}{F\!\!\!\!/\ }


\renewcommand{\epsilon}{\varepsilon}


\def\comment#1{}
\def\fixit#1{}

\def\equno#1{(\ref{#1})}
\def\sectno#1{section~\ref{#1}}
\def\figno#1{Fig.~(\ref{#1})}

\def\ellipsis{~$\ldots$ }
\def\D#1#2{{\partial #1 \over \partial #2}}
\def\tf#1#2{{\textstyle{#1 \over #2}}}   
\def\df#1#2{{\displaystyle{#1 \over #2}}}      
\def\eps{\epsilon}
\def\+{^\dagger}
\def\d{{\rm d}}
\def\e{{\rm e}}
\def\i{{\rm i}}
\def\diag{{\rm diag}}
\def\reg{{\rm regular}}
\def\t{\tilde}
\def\zb{\bar{z}}
\def\dim{{\rm dim}\,}
\def\unitdisk{\Delta}
\def\halfplane{{\bf H}}
\def\Complex{{\bf C}}
\def\sgn{{\rm sgn}\,}
\def\tr{{\rm tr}\,}
\def\binomial#1#2{\left( #1 \atop #2 \right)}

\section{Introduction}
\label{Intro}

Recent exciting developments in string theory suggest that it has
a variety of non-perturbative symmetries 
\cite{filq,sen,schwarz,Schwarz,duff,chpt,witten,strominger,jhas}. 
In general these duality symmetries
interchange fundamental strings with so-called $p$-branes,
objects with $p$ intrinsic dimensions. In a weakly coupled theory
the $p$-branes are heavy objects which may be thought of as
solitonic solutions to the equations describing 
the fundamental strings \cite{dh,mdjl,chs,hs}.
When extrapolated to strong coupling, however, the $p$-branes
become light and should therefore be regarded as the fundamental
degrees of freedom in the strongly coupled theory.

An unusual feature of the $p$-branes we are discussing is that their
masses scale as $1/g$, where $g$ is the string coupling, rather than
the familiar $1/g^2$ found for solitons in field theories.
This unusual dependence on $g$ follows from{} the fact that
the $p$-branes are stabilized by their R-R charges, and the R-R sector
of string theory couples to the dilaton differently from{} the NS-NS
sector \cite{witten}. The absence of the ``sphere term'' in the $p$-brane mass
suggests that it comes from{} the disk topology. This reveals a subtle 
connection of $p$-branes with open strings, even in the type~II
theory which has closed string excitations only. We will, of course, make
use of this relation to open strings in our calculations.

In the weak coupling limit the low-energy effective field theory
description of strings is well known, and its various $p$-brane
solutions were found years ago \cite{dh,mdjl,chs,hs}.
Because the importance of 
$p$-branes in the non-perturbative dynamics has become certain,
it became important to advance their understanding beyond the
narrow confines of the low-energy supergravity. A breakthrough
in this direction took place a few months ago when, building on earlier
work \cite{dlp,Green}, Polchinski showed 
\cite{polchinski} that the Dirichlet-branes (D-branes) 
carry precisely the
values of the R-R charges needed for the string duality symmetries.
Polchinski's work strongly suggests that the D-branes are the
exact stringy versions of the extremal black $p$-brane solutions
of the effective supergravity. The conformal field theory description of
the D-branes is extremely simple: one introduces auxiliary open strings
with Dirichlet boundary conditions in some directions replacing
the conventional Neumann boundary 
conditions \cite{dlp,Green}. Thus, these open strings
are not free to propagate everywhere, but are instead glued to
a $p+1$ dimensional hyperplane which is the world history of
the $p$-brane. This is how the hidden connection with open strings
alluded to previously is accomplished in type~II theory.
A nice physical picture for this, suggested by Witten \cite{witsem},
is that the auxiliary open string
is in reality a closed string, part of which is hidden behind a black
$p$-brane horizon.

The simplicity of the D-brane approach to the R-R charged string
solitons has originated a flurry of activity. Many $p$-brane
properties that were anticipated based on supersymmetry and duality
have been confirmed by explicit D-brane calculations
\cite{ed,witpol,li,ck}.  Furthermore, D-branes have made it possible
to formulate new conjectures and concepts
\cite{witinst,doug,vafa,bsv,horw,as,schmid}.  The stringy construction
of $p$-branes has made it possible to study those properties that do
not follow simply from{} BPS saturation arguments. A few such studies of
D-brane dynamics have appeared in the literature \cite{ck,kt,bachas},
and many more are undoubtedly possible. One interesting physical
question that was studied early on \cite{kt,bachas} concerns the size
of $p$-branes. The effective field theory has little to say about
this, but luckily D-branes do provide some stringy information.  For
instance, they may be probed by scattering of massless string states.
In \cite{kt} the lowest order scattering amplitudes for massless NS-NS
states off D-branes were calculated explicitly. The resulting formulae
revealed that, although classically all $p$-branes appear point-like,
the stringy quantum effects endow all the D-branes with $p>-1$ with a
size of order the string scale, $\sqrt{\alpha'}$. Furthermore, the
effective size grows indefinitely with increasing energy of the probe,
exhibiting the Regge behavior well-known in the high-energy scattering
of fundamental strings \cite{regge}. This suggests that an
instantaneous ``snapshot'' of a $p$-brane will show it smeared all
over space, similar to the ``snapshots'' of fundamental strings
\cite{kks}.

While the graviton scattering off D-branes revealed that their
gravitational radius exhibits the Regge behavior, it was not
explicitly shown that this also holds for the R-R charge radius.
In this paper we continue the program started in \cite{kt}
by calculating
the leading order scattering amplitude for two massless R-R states 
off D-branes.\footnote{In scattering off
D-branes we have also found non-vanishing two-point
functions mixing one R-R and one NS-NS massless states. 
We do not discuss them in this paper, but hope to present their
calculations in the near future.}
While the calculations with the R-R vertex
operators {\it a priori} seem technically challenging, we were
suprised to find that they are, in some ways, simpler than the corresponding
NS-NS calculations. We obtain a compact general formula for
the scattering of any R-R $n$-form field off any Dirichlet $p$-brane.
This formula explicitly shows that the Regge behavior and the exponential
fall-off at fixed angle found in NS-NS scattering also hold
for R-R scattering. The scattering amplitude exhibits an infinite
series of $t$-channel poles corresponding to exchange of closed string
states with the $p$-brane. The pole at $t=0$ corresponds to 
graviton and dilaton exchange and should therefore agree with the
amplitude found in the effective field theory. To check this, we study
the propagation of the $n$-form fields in the backgrounds of the extremal
black $p$-brane solutions to the type~II supergravity equations \cite{hs}.
We find that the residue of the $1/t$ pole in these field theoretic 
calculations matches precisely with that found in the stringy
D-brane calculations. This interesting result provides a good check
on the polarization dependence of the scattering off the D-branes
and gives a physical explanation of many of its 
features.\footnote{Our check on polarization dependences is similar in spirit
to the work in \cite{cmp}
where it was checked that appropriate generalizations
of the Dabholkar-Harvey solitons \cite{dh} approximately
describe long fundamental strings.}
More importantly, it strengthens the claim that the D-branes
are the stringy objects which, at large distances,
are approximated by the extremal black $p$-branes
of the effective supergravity.

The D-instanton ($-1$-brane) deserves special discussion because it
appears point-like even in the full stringy calculations 
\cite{Green,kt,jp2,mgpw,Green1,shenker}: instead
of an infinite sequence of $t$-channel poles there is only one pole
at $t=0$. To gain a better understanding of this phenomenon, 
we study a solution of the type~IIB supergravity equations which
describes an instanton carrying R-R ``charge'' and verify that
the field theoretic scattering off the D-instanton exactly reproduces
the full string amplitude. It is remarkable to find an object in
string theory that seems to be exactly described by the effective
field theory.

The organization of the paper is as follows.  In \sectno{Review} we briefly
summarize some of the amplitudes found previously for NS-NS scattering off
of $p$-branes.  In \sectno{TwoPoint} we present our main technical result,
namely the evaluation of all R-R scattering amplitudes off of $p$-branes.
In \sectno{Black} we show how the low momentum transfer limit of the
results of the previous two sections can be reproduced via scattering off
of the classical field of extremal black $p$-branes.  We conclude in 
\sectno{Discuss} with a discussion of how the general form of the
scattering amplitudes elucidates the relation of D-branes to black holes.

\section{Review of NS-NS scattering off D-branes}
\label{Review}

In \cite{kt}, D-branes were probed by scattering of gravitons and 
NS-NS antisymmetric tensor particles. For simplicity, their polarization 
tensors were taken to be non-vanishing only in the directions perpendicular
to the $p+1$ dimensional world volume, i.e.{} the following conditions were 
imposed:
\begin{equation}
\epsilon^{(1)}_{AB} = \epsilon^{(1)}_{Ai} =
\epsilon^{(2)}_{AB} = \epsilon^{(2)}_{Ai} = 0.
\end{equation}
Our conventions are to let capital indices run over the 
longitudinal directions, $A = 0, \ldots, p$, while lower-case indices 
run over the transverse directions, $i = p+1, \ldots, 9$, and 
Greek indices refer to all ten directions.
The graviton two-point function on the disk was found to be
\begin{equation}
F_g=-\epsilon^{(1)}_{ij}\epsilon^{(2)}_{ij}k_\parallel^2
{\Gamma (1+2k_\parallel^2) \Gamma (k\cdot p )
\over \Gamma (k\cdot p+2 k_\parallel^2+1)}\prod_{A=0}^p \delta (k^A + p^A)\ ,
\label{gravf}\end{equation}
where $k$ and $p$ are the momenta of the gravitons, and
\begin{equation}
k_\parallel^2 = k_A k^A= p_A p^A.
\end{equation}
The delta functions indicate that the momentum along the world history
is conserved.  The corresponding calculation for two anti-symmetric
tensor particles gave the following result:
\begin{equation}
F_B=\bigg [ 2 \epsilon^{(1)}_{ij}\epsilon^{(2)}_{il} p^j k^l 
-\epsilon^{(1)}_{ij}\epsilon^{(2)}_{ij}k_l p_l\bigg ]
{\Gamma (1+2k_\parallel^2) \Gamma (k\cdot p )
\over \Gamma (k\cdot p+2 k_\parallel^2+1)}\prod_{A=0}^p \delta (k^A + p^A).
\label{genant}\end{equation}
These formulae apply both to the ten-dimensional theory and to its
toroidal compactifications, where the $p$-brane may be wrapped around
some of the cycles.\footnote{In the compactified theories one may also
consider scattering of winding states, for which the results are
different from{} (\ref{gravf}), (\ref{genant}), but are also
explicitly computable.}  In the latter case, the momenta in the
compact directions are, of course, quantized, and the corresponding
Dirac delta functions are understood to be replaced by Kr\" onecker
delta functions.

The manifestly gauge invariant form of (\ref{genant}) is
\begin{equation}
F_B= 
{1\over 3} H^{(1)}_{ijl} H^{(2)}_{ijl} 
{\Gamma (1+2k_\parallel^2) \Gamma (k\cdot p )
\over \Gamma (k\cdot p+2 k_\parallel^2+1)}\prod_{A=0}^p \delta (k^A + p^A)\ ,
\label{gaugeant}\end{equation}
where
\begin{equation}
H^{(1)}_{\alpha \beta \gamma} = \i ( k_\alpha \epsilon^{(1)}_{\beta \gamma}
+ k_\beta \epsilon^{(1)}_{\gamma \alpha}+ k_\gamma \epsilon^{(1)}_{\alpha \beta})
\ .\end{equation}
Note that the term containing $H^{(1)}_{Ajl} H^{(2)}_{Ajl}$,
which could contribute to the amplitude for transverse polarizations,
is ``mysteriously'' absent. In \sectno{Black} this feature of 
the polarization
dependence will be reproduced for scattering off extremal
$p$-branes in the low-energy effective field theory.
For R-R scattering we will also find that field theory provides
a valuable check on the polarization dependence of the exact
string amplitude. This check will also strengthen our confidence in
the fact that the $p$-brane solutions in supergravity are simply
low-energy approximations to D-branes.

The D-instanton ($-1$-brane) is a special case where the low-energy
approximation becomes exact. For $p>-1$, (\ref{gravf}) and 
(\ref{genant}) exhibit an infinite sequence of poles in the $t$-channel,
and correspondingly in the $s$-channel. These characteristically stringy
properties of the scattering amplitudes
guarantee the Regge behavior and the exponential decrease with energy
at a fixed angle \cite{kt}. For $p=-1$, however, these stringy features
disappear because $k_\parallel^2=0$. The graviton two-point function
now vanishes, while that for the antisymmetric tensor simplifies down to 
\begin{equation}
{1\over 3 k\cdot p} H^{(1)}_{\alpha\beta\gamma} 
H^{(2)}_{\alpha\beta\gamma}
\label{instant}\end{equation}
In \sectno{Black} we will be able to reproduce this answer {\it exactly}
by scattering antisymmetric tensor particles off of the instanton
solution in supergravity.

\section{R-R scattering in string theory}
\label{TwoPoint}

This will be entirely devoted to world-sheet computations, so we begin by
reviewing some basic facts about the vertex operators for the massless
states in the R-R sector.  The vertex operator in the canonical ghost
picture for an R-R gauge boson with an $m$-form field strength polarization
$F_{(m)}$ and momentum $k$ is \cite{polchinski,dima}

\begin{equation}
\begin{array}{ll}
V(z,\zb) = 
   : \e^{-{1 \over 2} \phi(z)} S_\alpha(z) \e^{\i k \cdot X(z)} :
   \fslash_{(m)}^{\ \alpha}{}_\beta
   : \e^{-{1 \over 2} \t\phi(z)} \t{S}^\beta(\zb) \e^{\i k \cdot \t{X}(\zb)} :
   & {\rm (type\ IIA)} \\
V(z,\zb) =
   : \e^{-{1 \over 2} \phi(z)} S_\alpha(z) \e^{\i k \cdot X(z)} :
   \fslash_{(m)}^{\alpha\beta}
   : \e^{-{1 \over 2} \t\phi(z)} \t{S}_\beta(\zb) \e^{\i k \cdot \t{X}(\zb)} :
   & {\rm (type\ IIB)}
\end{array}                                               \label{VertexOp}
\end{equation}

\noindent
where we have defined

\begin{equation}
\fslash_{(m)} = \df{1}{m!} F_{\mu_1 \ldots \mu_m} 
   \gamma^{\mu_1} \cdots \gamma^{\mu_m} \ .           \label{GammaDef}
\end{equation}

\noindent
We will use a representation in which the $32 \times 32$ Dirac gamma
matrices are off-diagonal:

\begin{equation}
\gamma^\mu = \left( \begin{array}{cc}
                           0 & \gamma^{\mu\alpha\beta} \\
                           \gamma^\mu_{\alpha\beta} & 0
                    \end{array} \right) \ .
\end{equation}

\noindent
We normalize the $\gamma^\mu$ so that 
$\{\gamma^\mu,\gamma^\nu\} = -2 \eta^{\mu\nu}$, and we pick our
representation so that

\begin{equation}
\gamma_{11} = \gamma^0 \cdots \gamma^9 = 
              \left( \begin{array}{cc}
                              1 &  0 \\
                              0 & -1 
                     \end{array} \right) \ .
\end{equation}

\noindent
The leading order contribution to the scattering of an R-R boson off a 
$p$-brane, as illustrated in Fig.~1, comes from{} the two point 
function on the disk.\footnote{Similar disk calculations
with the conventional Neumann boundary conditions, and their comparison
with field theory, were attempted a long time ago \cite{fks}.}
Suppose the incoming
and outgoing bosons have field strength polarizations $F^{(1)}_{(m)}$ and 
$F^{(2)}_{(n)}$ and momenta $k$ and $-p$.  Then the amplitude is

\begin{figure}[t]
\centerline{\psfig{figure=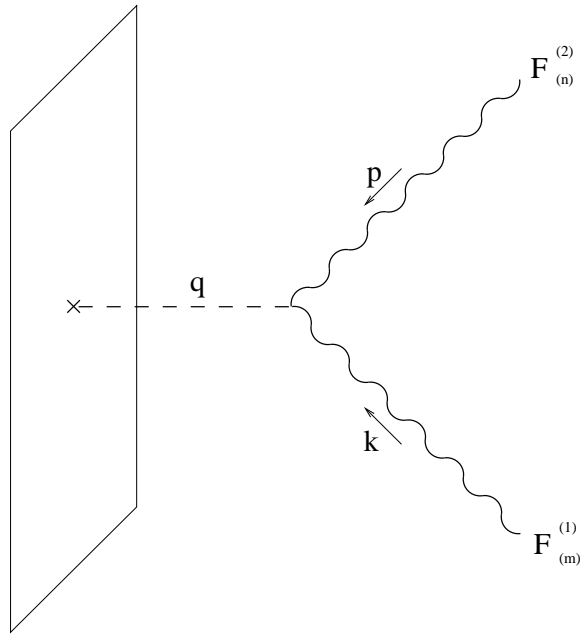,width=3in}}
\caption{Scattering of R-R gauge bosons off a $p$-brane.  The low momentum 
transfer limit is mediated as shown by $t$-channel exchange of a massless 
mode. }      \label{TExchange}
\end{figure}

\begin{equation}
F = \int_\unitdisk \d^2 z \, \left\langle V^{(1)}(z,\zb)
  V^{(2)}(0,0) \right\rangle \ .             \label{DiskAmp}
\end{equation}

\noindent
The computation of $F$ is possible in the canonical ghost picture because
the constraint on the superghost background charge is that the total sum of
the charges over both right and left moving sectors is $-2$.
Because of the way we will treat spin fields, it is convenient to
evaluate the two point function on the half-plane $\halfplane$ and then
conformally map the result to the disk to obtain the integrand of
\equno{DiskAmp}.

A convenient method for dealing with the mixed boundary conditions is 
to extend the holomorphic fields to the whole complex plane,
$\Complex$, in such a way that
their OPE's on $\Complex$ reproduce all the OPEs among holomorphic and 
antiholomorphic fields on $\halfplane$.  For example, to obtain Dirichlet
boundary conditions on all of the $\psi^\mu$ and $\t\psi^\mu$, we extend 
$\psi^\mu$ to $\Complex$ so that

\begin{equation}
\t\psi^\mu(\zb) = -\psi^\mu(\zb) \qquad {\rm for} \ z \in \halfplane \ .
                                                         \label{ExtendPsi}
\end{equation}

\noindent
Then the OPE

\begin{equation}
\psi^\mu(z) \psi^\nu(w) = \df{-\eta^{\mu\nu}}{z-w} + \reg
\end{equation}

\noindent
reproduces the desired OPEs for $\psi^\mu$ and $\t\psi^\mu$, the interesting
one of which is

\begin{equation}
\psi^\mu(z) \t\psi^\nu(\bar{w}) = \df{\eta^{\mu\nu}}{z-\bar{w}} + \reg \ .
\end{equation}

Given the boundary conditions appropriate for a $p$-brane, that is 
Neumann boundary conditions for $\mu = 0,\ldots,p$ and Dirichlet boundary
conditions for $\mu = p+1,\ldots,9$, it is clear (from{} bosonization, for 
example) that the analog of \equno{ExtendPsi} for spin fields is of the form

\begin{equation}
\begin{array}{ll}
\t{S}^\alpha(\zb) = M^{\alpha\beta} S_\beta(\zb) & {\rm (type\ IIA)}  \\
\t{S}_\alpha(\zb) = M_\alpha{}^\beta S_\beta(\zb) & {\rm (type\ IIB)} \ .
\end{array}                                          \label{ExtendS}
\end{equation}

\noindent
Determining the matrix $M$ seems somewhat nontrivial, but 
the solution is almost inevitable given the symmetries of the problem:

\begin{equation}
M = \gamma^0 \ldots \gamma^p                         \label{ResultM}
\end{equation}

\noindent
up to an overall power of $\i$ which is immaterial for our calculations.
To see how this result comes about for $1$-branes, note that if we bosonize
using

\begin{equation}
\df{\i}{\sqrt{2}} \left( \psi^{2i-1}(z) \pm  \i \psi^{2i}(z) \right) = 
  : \e^{\pm \i \phi_i(z)} :
\end{equation}

\noindent
where $\psi^{10} = \i \psi^0$, then the desired boundary conditions on
the $\psi^\mu$ are reproduced if we set

\begin{eqnarray}
\t\phi_i(\zb) &=& \phi_i(\zb) + \pi \quad {\rm for} \ i = 1,\ldots,4  
    \nonumber \\
\t\phi_5(\zb) &=& \phi_5(\zb) \ .
\end{eqnarray}

\noindent
Now it is easy to see that

\begin{equation}
\t{S}_\alpha(\zb) = : \e^{\i \lambda_{\alpha i} \t\phi_i(\zb)} :
  = (\sgn \lambda_{\alpha5}) S_\alpha(\zb) 
  = (\gamma^0 \gamma^1)_\alpha{}^\beta S_\beta(\zb) 
\end{equation}

\noindent
in this  representation where $\gamma^0 \gamma^1$ is diagonal.  Similar 
arguments go through in the case of general $p$.

Using \equno{ResultM} we can bring the two point function into a manageable
form: for type~IIB,

\begin{eqnarray}
\langle V^{(1)}(z,\zb) V^{(2)}(w,\bar{w}) \rangle &=& 
  (z-\zb)^{-1/4 + 2 k_\parallel^2} (w-\bar{w})^{-1/4 + 2 p_\parallel^2} 
          \nonumber \\
& & \ \ \cdot |z-w|^{2 (-1/4 + k \cdot p)} 
  |z-\bar{w}|^{2 (-1/4 + k_\parallel \cdot p_\parallel - 
   k_\perp \cdot p_\perp)} 
     \nonumber \\
& & \ \ \cdot \fslash_{(m)}^{(1)~\alpha\rho} M_\rho{}^\beta 
  \fslash_{(n)}^{(2)~\gamma\sigma} M_\sigma{}^\delta
  \langle S_\alpha(z) S_\beta(\zb) S_\gamma(w) S_\delta(\bar{w}) \rangle 
     \nonumber \\
& & \ \ \cdot \prod_{A=0}^p \delta (k^A + p^A)\ .
                                                   \label{CombVV}
\end{eqnarray}

\noindent
The four point function of spin fields has been calculated \cite{FMS}, and 
for $\gamma^\mu_{\alpha\beta}$ symmetric in $\alpha,\beta$ it reads

\begin{equation}
\langle S_\alpha(z_1) S_\beta(z_2) S_\gamma(z_3) S_\delta(z_4)\rangle = 
  \df{z_{14} z_{23} \gamma^\mu_{\alpha\beta} \gamma_{\mu\gamma\delta} - 
      z_{12} z_{34} \gamma^\mu_{\alpha\delta} \gamma_{\mu\beta\gamma} }{
      2 (z_{12} z_{13} z_{14} z_{23} z_{24} z_{34})^{3/4} }   \label{FourS}
\end{equation}

\noindent 
(the factor of $2$ in the denominator compensates for our
normalization of the $\gamma^\mu$, which is different from{}
\cite{FMS}).  The final result for the two point function on the half
plane is

\begin{eqnarray}
\langle V^{(1)}(z,\zb) V^{(2)}(w,\bar{w}) \rangle &=&
   (z-\zb)^{-1 + 2 k_\parallel^2} (w-\bar{w})^{-1 + 2 p_\parallel^2} 
       \nonumber \\
& & \ \ \cdot |z-w|^{2 (-1 + k \cdot p)}
  |z-\bar{w}|^{2 (-1 + k_\parallel \cdot p_\parallel - k_\perp \cdot p_\perp)}
       \nonumber \\
& & \ \ \cdot \tf{1}{2} \left( |z-\bar{w}|^2 P_1 - (z-\zb) (w-\bar{w}) P_2 
     \right)  \nonumber \\
& & \ \ \cdot \prod_{A=0}^p \delta (k^A + p^A)               \label{FinalVV}
\end{eqnarray}

\noindent
where 

\begin{eqnarray}
P_1 &=& \left( \fslash_{(m)}^{(1)~\alpha\rho} M_\rho{}^\beta 
               \gamma^\mu_{\beta_\alpha} \right)
        \left( \fslash_{(n)}^{(2)~\gamma\sigma} M_\sigma{}^\delta
               \gamma_{\mu\delta\gamma} \right)  
    = \tr \left( \df{1+\gamma_{11}}{2} \fslash_{(m)}^{(1)} M \gamma^\mu \right) 
      \tr \left( \df{1+\gamma_{11}}{2} \fslash_{(n)}^{(2)} M \gamma_\mu \right)  
          \nonumber \\
P_2 &=& \fslash_{(m)}^{(1)~\alpha\rho} M_\rho{}^\beta
               \gamma^\mu_{\beta_\gamma}
               \fslash_{(n)}^{(2)~\gamma\sigma} M_\sigma{}^\delta
               \gamma_{\mu\delta\alpha} 
    = \tr \left( \df{1+\gamma_{11}}{2} \fslash_{(m)}^{(1)} M \gamma^\mu 
                                       \fslash_{(n)}^{(2)}
 M \gamma_\mu \right) \ .
                                                     \label{P1P2}
\end{eqnarray}

\noindent
Writing $P_1$ and $P_2$ in terms of $32\times 32$ gamma matrices 
rather than $16\times 16$ blocks is convenient because these expressions
are correct both for type~IIA and type~IIB, whereas the index structure for 
the $16\times 16$ block expressions is different in the two cases.
The factors of $(1-\gamma_{11})/2$ enforce the sums over the correct type of
indices, upper or lower.

Mapping back to the unit disk and setting $x = r^2$, where
$r$ is the radial coordinate, we find that the 
scattering amplitude is 

\begin{eqnarray}
F &=& \int_0^1 \d x \, (1-x)^{-1 + 2 k_\parallel^2} x^{-1 + k \cdot p} 
   \left( P_1 + (1-x) P_2 \right) \prod_{A=0}^p \delta (k^A + p^A)
  \nonumber \\
  &=& \df{\Gamma(1 + 2 k_\parallel^2) \Gamma(k \cdot p) }{ 
          \Gamma(1 + 2 k_\parallel^2 + k \cdot p) }
      \left[ \df{2 k_\parallel^2 + k \cdot p }{ 2 k_\parallel^2} P_1 + 
       P_2 \right] \prod_{A=0}^p \delta (k^A + p^A) \nonumber \\
  &=& \df{\Gamma(1-2s) \Gamma(-t/2) }{ \Gamma(1-2s-t/2)} 
      \left[ \df{s+t/4}{s} P_1 + P_2 \right]
   \prod_{A=0}^p \delta (k^A + p^A)\ ,
                                                      \label{FinalF}
\end{eqnarray}

\noindent
where we have defined $s = -k_\parallel^2$ and $t = -(p+k)^2$.  This
is our main result: the two-point function of any R-R $n$-form fields
off of a Dirichlet $p$-brane is reduced to evaluating traces of gamma
matrices in ten dimensions.\footnote{Equation \equno{FinalF} is
correct only up to an overall factor, which includes not only the
trivial factors of $2$ and $\pi$ which we have neglected, but also
factors related to the overall normalization of disk amplitudes as
compared to sphere amplitudes.  These latter factors are necessary in
order to make a completely rigorous comparison with field theory based
on the effective action for type~II theories.  We will be satisfied in
the following section to demonstrate agreement in the infrared between
string theory and field theory up to such overall factors.}  The
dependence on polarizations is contained in $P_1$ and $P_2$, and all
amplitudes are multiplied by the universal prefactor
\begin{equation}
{\Gamma (1+2k_\parallel^2) \Gamma (k\cdot p )
\over \Gamma (k\cdot p+ 2 k_\parallel^2+1)}.
\end{equation}
This prefactor is the same as in the NS-NS scattering amplitudes,
(\ref{gravf})-(\ref{gaugeant}). For $p>-1$ this prefactor contains
an infinite series of poles and thus guarantees the Regge
behavior of D-branes. For $p=-1$,
$k_\parallel^2=0$ and the prefactor collapses to a single
pole, $1/(p\cdot k)$, which implies a field theoretic structure of
the scattering in the D-instanton background
(the amplitude, in fact, diverges if $P_1\neq 0$, which
is the case for the scattering of
R-R scalars only). In summary, as anticipated
in \cite{kt} all the qualitative features of the D-branes found from{}
NS-NS scattering apply to R-R scattering as well.

The traces in \equno{P1P2} are straightforward but tedious to evaluate.
The main tool one uses is 
the general (anti)-commutator of anti-symmetrized gamma matrices:

\begin{eqnarray}
\lefteqn{\left[ \gamma^{[\mu_1} \cdots \gamma^{\mu_m]},
       \gamma_{[\nu_1} \cdots \gamma_{\nu_n]} \right]_{(-1)^{mn+1}} =}
  \nonumber \\
 & & \sum_{j=1}^m (-1)^{1+mj+j(j+1)/2} \binomial{m}{j} \binomial{n}{j} 
      2^j j! \, \delta^{[\mu_1}_{[\nu_1} \cdots \delta^{\mu_j}_{\nu_j} 
      \gamma^{\mu_{j+1}} \cdots \gamma^{\mu_m]} 
      \gamma_{\nu_{j+1}} \cdots \gamma_{\nu_n]} \ ,
\end{eqnarray}

\noindent
which is essentially Wick's theorem for gamma matrices.
The result is

\begin{eqnarray}
P_1 &=& -256 (p+2) \Big\langle \delta_{m-p-2} F^{(1)}_{(m)} + 
         (-1)^{m(m+1)/2} \delta_{m+p-8} *F^{(1)}_{(m)},\,  \nonumber \\
    & & \qquad\qquad\qquad\ 
         \delta_{n-p-2} F^{(2)}_{(n)} +
         (-1)^{n(n+1)/2} \delta_{n+p-8} *F^{(2)}_{(n)} 
        \Big\rangle^{p+2}_{p+1}  \nonumber \\
    & & \ \ -256 \Big\langle \delta_{m-p} F^{(1)}_{(m)} + 
         (-1)^{m(m+1)/2} \delta_{m+p-10} *F^{(1)}_{(m)},\,  \nonumber \\
    & & \qquad\quad\ 
         \delta_{n-p} F^{(2)}_{(n)} +
         (-1)^{n(n+1)/2} \delta_{n+p-10} *F^{(2)}_{(n)}
        \Big\rangle^p_p  \nonumber \\
P_2 &=& (-1)^{mp + m(m+1)/2 + p(p+1)/2} \, 
         32 \sum_{j=0}^{{\rm min}\, (p+1,n)}
          (-1)^j 2^j (4-n-p+j) \binomial{n}{j}  \nonumber \\
    & & \ \ \cdot \left\langle \delta_{m-n} F^{(1)}_{(m)} + 
           (-1)^{m(m+1)/2} \delta_{m+n-10} *F^{(1)}_{(m)},\, F^{(2)}_{(n)}
          \right\rangle^n_j \ .                          \label{FinalP1P2}
\end{eqnarray}

\noindent
The notation requires some explanation.  If $U_{(n)}$ and $V_{(n)}$ are
$n$-forms, then 

\begin{equation}
\left\langle U_{(n)},\, V_{(n)} \right\rangle^n_j = 
 \df{1}{n!} \eta^{A_1 B_1} \cdots \eta^{A_j B_j} 
            \eta^{\mu_{j+1} \nu_{j+1}} \cdots \eta^{\mu_n \nu_n}
          U_{A_1 \ldots A_j \mu_{j+1} \ldots \mu_n}
          V_{B_1 \ldots B_j \nu_{j+1} \ldots \nu_n}       \label{BracketDef}
\end{equation}

\noindent
where as usual the $A_i$ and $B_i$ run only from{}~$0$ to~$p$ while the 
$\mu_i$ and $\nu_i$ from{}~$0$ to~$9$.  Hodge duals are defined by

\begin{equation}
(*U_{(n)})_{\mu_1 \ldots \mu_{10-n}} = 
  \df{1}{n!} \eps_{\mu_1 \ldots \mu_{10}} U^{\mu_{11-n} \ldots \mu_{10}}  \ .
\end{equation}

\noindent
In the following section we write out the polarization dependences 
explicitly for some special cases and find a simple physical meaning of
these formulae.

\section{Scattering off of black $p$-branes}
\label{Black}

Now, as promised, we will indicate how the $t=0$ pole of the scattering
amplitudes \equno{FinalF} can be reproduced by the low-energy effective field
theory.  For $p>-1$, the $t=0$ pole is

\begin{equation}
F_{\rm string\ theory} = -\df{2}{t} (P_1 + P_2)         \label{StringPole}
\end{equation}

\noindent 
In this equation and in the rest of this section, we suppress the 
momentum conserving delta functions in \equno{FinalF}.  
In principle, one should be able to reproduce \equno{StringPole}
by directly evaluating the graphs corresponding to the $t$-channel
exchange of a massless particle, as in Fig.~1.  However, it
is simpler to consider an equivalent description of the phenomena:
scattering at large impact parameter of R-R bosons by the classical 
gravitational field of 
extremal black $p$-branes.  To put it in astrophysical terms,
we want to look at gravitational lensing of R-R light by $p$-branes.

Because $p$-branes are BPS saturated, we expect their infrared properties
to be described by extreme R-R charged $p$-brane solutions to the 
low-energy effective action:  

\begin{eqnarray}
\d s^2 &=& A^{-1/2} \left( -\d t^2 + \d x_1^2 + \ldots + \d x_p^2 \right) + 
           A^{1/2} \left( \d y^2 + y^2 \d \Omega_{8-p}^2 \right)  \nonumber \\
\e^{-2 \phi} &=& A^{(p-3)/2}  \nonumber \\
F_{(p+2)} &=& \df{Q}{y^{8-p}} A^{-2} \d t \wedge \d x_1 \wedge \ldots 
   \wedge \d x_p \wedge \d y                             \label{SolutionP}
\end{eqnarray}

\noindent
where $F_{(p+2)}$ is the R-R field strength coupling to the brane, and 

\begin{equation}
A = 1 + \df{2}{7-p} \df{Q}{y^{7-p}} \ .                  \label{DefineA}
\end{equation}

\noindent
For $p>-1$ these solutions may be written down simply by a coordinate
transformation of the extremal R-R charged $p$-branes found in
\cite{hs}\footnote{As explained in \cite{hs}, the 3-brane is a special
case because it couples to the self-dual 5-form field strength;
therefore, the third line of (\ref{SolutionP}) is modified.}.  While
\cite{hs} discussed only $0\leq p\leq 6$, solutions (\ref{SolutionP}),
(\ref{DefineA}) may be extrapolated in an obvious way to $p=7$ \cite{ggp}
and $p=8$.  For the $7$-brane the solution is (\ref{SolutionP}) with
\begin{equation}
A = 1 + 2 Q\ln(y/y_0) \ ,                  
\end{equation}
while for the $8$-brane,
\begin{equation}
A = 1 + 2 Qy= 1+ 2Q|x_9| \ .                  
\end{equation}
A new feature we find for $p=7$ and $8$ is that $A$ grows with the distance
from{} the $p$-brane. Thus, the geometry is not asymptotically flat.

The D-instanton ($p=-1$) is a special case
which we are going to discuss in some detail.
The D-instanton is a source of the R-R scalar, which is 
essentially a ten-dimensional axion. In order to make its 
axionic properties manifest, we are going to use its dual, $8$-form,
description. Our goal, therefore, is to find an instanton solution
to the following euclidean action,
\begin{equation}
S= \int d^{10} x \sqrt G \biggl [e^{-2\phi} (-R+ 4 (\partial \phi)^2 )
+\df{1}{2\cdot 9!} F_{(9)}^2 \biggr ]
\end{equation}
It is not hard to verify that the following is a solution:\footnote{
After we independently found this solution we were informed by
M.~Green that it was originally obtained in \cite{ggp}.}
\begin{eqnarray}
\d s^2 &=&  A^{1/2} \left( \d y^2 + y^2 \d \Omega_9^2 \right)  \nonumber \\
\e^{-2 \phi} &=& A^{-2}  \nonumber \\
F_{(9)} &=& \df{Q}{y^9} A^{-2} * \d y                \label{SolutionD}
\end{eqnarray}
where 
\begin{equation}
A = 1 + \df{1}{4} \df{Q}{y^8} \ .                  \label{DefineAD}
\end{equation}
This solution has the following
interesting feature: the Einstein metric,
\begin{equation}
g_{ij} = G_{ij} e^{-\phi/2} = \delta_{ij}\ ,
\end{equation}
is flat!  Since the Einstein metric describes the physical gravitational
field, we conclude that the R-R charged instanton in 10 dimensions
emits a dilaton, but no gravitational field.

Large impact parameter scattering can be computed by expanding 
in powers of 

\begin{equation}
\df{1}{7-p} \df{Q}{y^{7-p}} = \df{2 \pi^{(9-p)/2} }{ \Gamma((9-p)/2) }
  \int \df{\d^{9-p} q}{(2 \pi)^{9-p}} \, \e^{\i q \cdot y} \df{Q}{q^2} \ .
                                                         \label{LaplaceKernel}
\end{equation}

\noindent
Let us consider scattering of $n$-form field strength bosons.  Because
the gauge fields are abelian, they have no self-interactions and
are therefore insensitive to their own background values.
The relevant piece of the effective action is

\begin{eqnarray}
S &=& \int \d^{10} x \, \sqrt{G} \df{1}{2 \cdot  n!} F_{(n)}^2  \nonumber \\
  &=& \int \d^{10} x \, \tf{1}{2} 
        \left\langle F_{(n)},F_{(n)} \right\rangle^n_0  \nonumber \\
  && + 
       \int \d^{10} x \, \tf{1}{2} \df{1}{7-p} \, \df{Q}{y^{7-p}} 
        \Big[ (4-p-n) \left\langle F_{(n)},F_{(n)} \right\rangle^n_0 
        + 2n \left\langle F_{(n)},F_{(n)} \right\rangle^n_1 \Big] \nonumber \\
& &  \qquad +   \ldots                                \label{SExpanded}
\end{eqnarray}
where in the second and third lines we have expanded up to the one
graviton contribution.  The angle bracket notation still indicates
contraction of indices with the metric $\eta^{\mu\nu}$, as in 
\equno{BracketDef}.  Setting $q = p+k$, 
one can see from{} the right hand sides of
\equno{LaplaceKernel} and \equno{SExpanded} that the single graviton
contribution determines the $1/t$ pole of the scattering amplitude.
Neglecting normalization factors as usual, the $1/t$ pole is

\begin{equation}
F_{\rm field\ theory}
  = \df{1}{t} \left[ (4-p-n) 
         \left\langle F^{(1)}_{(n)},F^{(2)}_{(n)} \right\rangle^n_0 + 
      2n \left\langle F^{(1)}_{(n)},F^{(2)}_{(n)} \right\rangle^n_1 \right] \ .
                                                        \label{FTPole}
\end{equation}

\noindent
The test of whether Dirichlet $p$-branes bend R-R light like 
the extremal black
$p$-branes (at low momentum transfer, of course) is to compare the
polarization dependence of
\equno{StringPole} and \equno{FTPole}.  
Let us look at some examples:

\begin{itemize}

\item $p=0$, $n=2$.  This is the most physically familiar case: 
gravitational lensing of vector gauge bosons by a black hole.  It is 
not hard to work out $P_1$ and $P_2$ directly from{} \equno{P1P2}:

\begin{eqnarray}
P_1 &=& -512 \left\langle F^{(1)}_{(2)},F^{(2)}_{(2)} \right\rangle^2_1 
           \nonumber \\
P_2 &=& -64 \left\langle F^{(1)}_{(2)},F^{(2)}_{(2)} \right\rangle^2_0 + 
      384 \left\langle F^{(1)}_{(2)},F^{(2)}_{(2)} \right\rangle^2_1  
           \nonumber \\
F_{\rm string\ theory} &=& 
 \df{128}{t} 
   \left[ \left\langle F^{(1)}_{(2)},F^{(2)}_{(2)} \right\rangle^2_0 + 
        2 \left\langle F^{(1)}_{(2)},F^{(2)}_{(2)} \right\rangle^2_1 \right]
\nonumber \\
&\sim & {1\over t} \left[F^{(1)}_{ij} F^{(2)}_{ij}+
4 F^{(1)}_{Aj} F^{(2)}_{Aj} + 3 F^{(1)}_{AB} F^{(2)}_{AB} \right].
\end{eqnarray}

\noindent
This matches the field theory result

\begin{equation}
F_{\rm field\ theory} = \df{2}{t} 
  \left[ \left\langle F^{(1)}_{(2)},F^{(2)}_{(2)} \right\rangle^2_0 +
       2 \left\langle F^{(1)}_{(2)},F^{(2)}_{(2)} \right\rangle^2_1 \right] \ .
\end{equation}

\item $p=1$, $n=3$.

\begin{eqnarray}
P_1 &=& 768 \left\langle F^{(1)}_{(3)},F^{(2)}_{(3)} \right\rangle^3_2  
          \nonumber \\
P_2 &=& 192 \left\langle F^{(1)}_{(3)},F^{(2)}_{(3)} \right\rangle^3_1 - 
        768 \left\langle F^{(1)}_{(3)},F^{(2)}_{(3)} \right\rangle^3_2  
          \nonumber \\
F_{\rm string\ theory} &=& -\df{384}{t} 
            \left\langle F^{(1)}_{(3)},F^{(2)}_{(3)} \right\rangle^3_1
\nonumber \\
&\sim & {1\over t} \left[F^{(1)}_{Aij} F^{(2)}_{Aij}+
2 F^{(1)}_{ABj} F^{(2)}_{ABj} + F^{(1)}_{ABC} F^{(2)}_{ABC} \right]
\label{poldep}\end{eqnarray}

again matching field theory:

\begin{equation}
F_{\rm field\ theory} = \df{6}{t} 
            \left\langle F^{(1)}_{(3)},F^{(2)}_{(3)} \right\rangle^3_1 \ .
\end{equation}

\item $p=4$, $n=4$. This is a more complex example:
scattering of a 4-form field strength off a 4-brane.
Here (\ref{StringPole}) yields

\begin{eqnarray}
F_{\rm string\ theory} &=& -\df{256}{t} \biggl [
            \left\langle F^{(1)}_{(4)},F^{(2)}_{(4)} \right\rangle^4_4
+ 6 \left\langle *F^{(1)}_{(4)},*F^{(2)}_{(4)} \right\rangle^6_5
\nonumber \\
&+& 3 \sum_{j=0}^3 {(-1)^j 2^j \over j! (3-j)! }
\left\langle F^{(1)}_{(4)},F^{(2)}_{(4)} \right\rangle^4_j \biggr ]
\nonumber \\
&\sim & {1\over t} \left[F^{(1)}_{ijkl} F^{(2)}_{ijkl}+
2F^{(1)}_{Aijk} F^{(2)}_{Aijk}-
2 F^{(1)}_{ABCj} F^{(2)}_{ABCj} - F^{(1)}_{ABCD} F^{(2)}_{ABCD} \right].
\end{eqnarray}

The field theory calculation gives

\begin{eqnarray}
F_{\rm field\ theory} &=& -\df{4}{t} 
  \left[ \left\langle F^{(1)}_{(4)},F^{(2)}_{(4)} \right\rangle^4_0 -
       2 \left\langle F^{(1)}_{(4)},F^{(2)}_{(4)} \right\rangle^4_1 \right]
\nonumber \\
&\sim & {1\over t} \left[F^{(1)}_{ijkl} F^{(2)}_{ijkl}+
2F^{(1)}_{Aijk} F^{(2)}_{Aijk}-
2 F^{(1)}_{ABCj} F^{(2)}_{ABCj} - F^{(1)}_{ABCD} F^{(2)}_{ABCD} \right],
\end{eqnarray}

\noindent
once again reproducing the correct relative factors between terms.
As this case exemplifies, terms of the form 
$\left\langle *F^{(1)}_{(n)},*F^{(2)}_{(n)} \right\rangle^{10-n}_j$ 
appear in the string theory expressions for $n+p = 8$~or~$10$.  The bracket 
notation then becomes inconvenient and it is necessary to
expand out all expressions into $SO(1,p) \times SO(9-p)$ invariants
$F^{(1)}_{A_1 \ldots A_l i_{l+1} \ldots i_n} 
 F^{(2)}_{A_1 \ldots A_l i_{l+1} \ldots i_n}$ 
before a comparison with field theory can be made.

\item $p=-1$.  In this case, the kinematics changes slightly: because 
there are no directions parallel to the brane, $s=0$.  As a result, 
the beta function prefactor simplifies.  For $n=3$, the full result 
\equno{FinalF} reduces to 

\begin{eqnarray}
F_{\rm string\ theory} 
  &=& \df{128}{t} 
    \left\langle F^{(1)}_{(n)}, F^{(2)}_{(n)} \right\rangle^n_0  \nonumber \\
  &\sim& \df{1}{t} F^{(1)}_{\mu_1 \ldots \mu_n} F^{(2)}_{\mu_1 \ldots \mu_n} 
                                                      \label{FSI}
\end{eqnarray}

\noindent
which is reproduced {\it exactly} by the one-graviton field theory
calculation using the solution \equno{SolutionD}:

\begin{equation}
F_{\rm field\ theory} = \df{2}{t} 
    \left\langle F^{(1)}_{(n)}, F^{(2)}_{(n)} \right\rangle^n_0   \label{FFI}
\end{equation}

For $n=1$, $P_1 \neq 0$, and so the amplitude \equno{FinalF} diverges.  
However, because of the double trace form of $P_1$, we suspect that the
divergent term is canceled by the disk--cylinder contribution, which is
of the same order in $g$. In this world sheet configuration
one R-R scalar vertex operator is inserted 
on a disk and another on a cylinder.\footnote{Such 
disconnected world sheets are well-known to contribute to
D-instanton processes \cite{jp2}. }
If the divergence cancels, then the string theory
and one-graviton field theory calculations again agree exactly.

\item $n=5$. We find that the scattering amplitudes
for the self-dual 5-forms off of any $p$-brane {\it vanish} both in
string theory and in field theory. Here one encounters an interesting
subtlety: 
it is impossible to write down a covariant action for the
self-dual $5$-form in type IIB supergravity. 
The procedure that seems to work, however,
is to calculate the two-point function from the standard 
quadratic action
for a general $5$-form field,
\begin{equation}
S= \int d^{10} x \sqrt G
\df{1}{2\cdot 5!} F_{(5)}^2
\ ,\end{equation}
and to impose the self-duality constraint on the answer.
This constraint makes the residue of the $t=0$ pole,
(\ref{FTPole}), vanish for all $p$.
\end{itemize}

Altogether there are $25$ different scattering processes where our field
theory analysis serves as a check on the string theory computation: 
$n=1,3,5$ for $p$ odd and $n=2,4$ for $p$ even, with $-1 \leq p \leq 8$.  
We have checked the agreement
between string theory and field theory in all  
of the $25$ possible cases,
some of them involving quite nontrivial cancelations between $P_1$
and $P_2$. 
The consistent agreement in the polarization dependence seems
to us strong evidence for regarding extreme black $p$-branes as the
low-energy description of Dirichlet $p$-branes.

Another interesting check on our results is provided by performing a
T-duality transformation on all of the coordinates.  This
transformation interchanges the Neumann with the Dirichlet boundary
conditions, which means that the $p+1$ directions along the brane
world volume are interchanged with the $9-p$ transverse directions.
Thus, the T-duality relates scattering of $n$-forms off $p$-branes to
scattering of $n$-forms off $8-p$-branes: one simply interchanges the
transverse (lower-case) with the longitudinal (capital) indices in the
polarization dependence. For instance, from{} the polarization
dependence (\ref{poldep}) for $p=1$, $n=3$ we infer the polarization
dependence for $p=7$, $n=3$:
\begin{equation}
\sim \left[F^{(1)}_{ijk} F^{(2)}_{ijk}+
2 F^{(1)}_{Aij} F^{(2)}_{Aij} + F^{(1)}_{ABi} F^{(2)}_{ABi} \right].
\end{equation}
This is easily confirmed by explicit calculations both in string
theory and in field theory.

In conclusion, let us note that the field theory also explains the
$1/t$ pole in the string NS-NS amplitudes. Consider, for instance, the
scattering of NS-NS antisymmetric tensors off a $p$-brane.  The
relevant part of the effective action is
\begin{equation}
\int \d^{10} x \, \sqrt{G}e^{-2\phi} \df{1}{12} H_{\mu\nu\lambda}^2  .
\end{equation}
Substituting, for instance, the $0$-brane solution we find that small
angle scattering is described by
\begin{equation}
{1\over t} H^{(1)}_{ijl} H^{(2)}_{ijl}.
\end{equation}
Just as in the string amplitude~(\ref{gaugeant}), the term containing
$H^{(1)}_{0jl} H^{(2)}_{0jl}$ is absent!  Similar checks of the
polarization dependence work out for the scattering of gravitons.

In the D-instanton case we find that the field theory amplitude
reproduces the complete string amplitude (\ref{instant}). This
illustrates the special point-like nature of this object.  The field
theory results are further valuable as a guide towards future string
calculations. For example, in scattering NS-NS antisymmetric tensors
with arbitrary polarizations off $p$-branes we find the following
universal formula:
\begin{equation}
{1\over t}\left [ H^{(1)}_{ijl} H^{(2)}_{ijl}
-3 H^{(1)}_{ABl} H^{(2)}_{ABl} -
2 H^{(1)}_{ABC} H^{(2)}_{ABC}\right ].
\end{equation}
This pole at $t=0$ should be reproduced by the string calculation
with arbitrary polarizations, which we leave for future work.

\section{Discussion}
\label{Discuss}

The low-energy supergravity includes massless fields only and is,
therefore, only capable of reproducing the pole at $t=0$ in the exact
scattering amplitude ($t=2p\cdot k$). The poles at negative integer values 
of $t$ come from{} the fact that the $p$-brane also excites an
infinite set of massive background fields. This could
be anticipated on general grounds.
One feature of the exact D-brane scattering that could hardly
be anticipated, however, is the infinite sequence of $s$-channel
poles at integer values of $\alpha' k_\parallel^2$.
These poles reveal a discrete spectrum of D-brane excitations associated
with the massive states of the auxiliary open strings. Since these
open strings are in reality closed strings partly trapped behind a 
horizon, they implement a stringy version of the horizon dynamics
anticipated by the ``stretched horizon'' ideas \cite{stu}. 
It is thus plausible
to identify $D$-branes containing such horizon excitations 
with non-extremal $p$-branes.

Let us review the particularly interesting example of the $0$-brane
in type~IIA theory,
which was discussed in detail in \cite{kt}. The unexcited Dirichlet $0$-brane
describes an extremal ten-dimensional dilatonic black hole, carrying
a basic unit of the R-R electric charge. Thus, the $0$-brane gives
us a new insight into black holes in $d=10$ (black holes in $d<10$
could be studied by wrapping fundamental strings around the cycles of
tori \cite{cmp}). The mass of the extremal black hole is
$M_0=1/(g\sqrt{\alpha'})$. 
Excited, non-extremal states are obtained by
attaching the auxiliary open strings to the $0$-brane.
The masses of such states are
\begin{equation}
M = {1\over g\sqrt{\alpha'}} + \sum_{i=1}^k\sqrt{n_i\over \alpha'}
+ O(g)
\ ,\end{equation}
where we have $k$ open strings with excitation levels $n_i$.
We believe that these excited states should be interpreted as
non-extremal black holes. Thus, the black hole spectrum is discrete,
but becomes continuous in the limit of high excitation numbers (we should
also keep in mind that all the $n_i>0$ states have finite widths).
It is remarkable that the gaps in the spectrum do not vanish even in
the weak coupling limit.\footnote{We are grateful to C. G. Callan for
pointing this out to us.} Perhaps this points to an 11-dimensional 
interpretation of the discrete spectrum (the connection of the 
type~IIA theory with some unknown theory in 11 dimensions was first 
proposed in \cite{chpt,witten}).

The degeneracy of black holes of mass $M$ is given by the number of
states of the open string gas with total energy $E=M-M_0$.  The number
of open string states at oscillator level $n$, $d_n$, is well known to
grow exponentially for large $n$:
\begin{equation}
d_n\sim n^{-11/4} \exp \left (\pi\sqrt{8n}\right ).
\end{equation}
This number includes states of spin $J$ ranging from{} $0$ to $n+1$.
If we restrict ourselves to non-rotating black holes, we need to
isolate the number of $J=0$ open string states at level $n$. This has
been calculated in \cite{russo} and turns out to be suppressed only by
a factor $1/\sqrt n$ compared to $d_n$:
\begin{equation}
d^{(J=0)}_n\sim n^{-13/4} \exp \left (\pi\sqrt{8n}\right ).
\end{equation}
If the total energy is
\begin{equation}
E=\sqrt{N\over \alpha'}= 
\sum_{i=1}^k\sqrt{n_i\over \alpha'}
\end{equation}
then the degeneracy is
\begin{equation}
d_{\{ n_i\} }=\prod_i d_{n_i}\sim d_N
\end{equation}
where we have ignored the slowly varying prefactors multiplying
the exponentials. Thus, each multiple-string configuration
contributes with roughly the same weight to the degeneracy.
This means that the degeneracy is $d_N$ times the number of ways the total
energy may be divided among different open strings. This number,
$\Omega_N$, grows exponentially, but slower than $d_N$,
\begin{equation}
\Omega_N\sim \exp \left (c N^{1/3} \right ).
\end{equation}
Thus, the entropy of the ten-dimensional black holes, which
are non-extremal excitations
of the extremal R-R charged black hole, is given by
\begin{equation}
S = \ln (\Omega_N d_N)=\pi\sqrt{ 8\alpha'} (M-M_0) + O
\biggl ( (M-M_0)^{2/3}  
\biggr ). \label{entro}
\end{equation}

It is interesting to compare the Dirichlet $0$-brane result with
the Bekenstein-Hawking entropy of the near-extremal RR-charged black hole
in 10 dimensions. The relevant supergravity solution was found in
\cite{hs},
\begin{eqnarray}
\d s^2 &=& -\Delta_+ \Delta_-^{-1/2}\d t^2 +  \Delta_+^{-1} \Delta_-^{-17/14}
           \d r^2 + r^2\Delta_-^{-3/14} \d \Omega_8^2 \nonumber \\
e^{-2\phi} &=& \Delta_-^{3/2}
\end{eqnarray}
where 
\begin{equation}
\Delta_\pm (r)  = 1 - {r_\pm^7\over r^7}
\end{equation}
Transforming the above to the Einstein metric, we find that
the ``area'' of the horizon is
\begin{equation}
A=\omega_8 r_+^8 \left [1 - {r_-^7\over r_+^7}\right ]^{9/14}
\end{equation}
where $\omega_8$ is the ``area'' of a unit sphere in 8 dimensions.
The ADM mass is given by
\begin{equation}
M={8 r_+^7 - r_-^7\over 2 \kappa^2}
\end{equation}
For nearly extremal black holes, $r_\pm= r_0\pm \delta r$. In this regime,
the Bekenstein Hawking entropy, $2\pi A/\kappa^2$, scales as
\begin{equation}
S_{BH} \sim M_0^{1/2} (\delta M)^{9/14}
\label{entroclass} \end{equation}
This is clearly different from the scaling $S\sim \delta M$ found using
the D-brane counting of states. Notice however that in equation
(\ref{entro}) we have assumed that we have just a single zero brane.
If we consider $n_0$ zero branes,
 the second term in (\ref{entro}) is  
more important and gives a contribution that, in the simplest
approximation (disregarding completely the first term),
  behaves as  $ S \sim M_0^{2/3 }(\delta M)^{2/3} $
which is indeed closer to  (\ref{entroclass}).  
The discrepancy may be attributed to
the fact that the string coupling is very strong near the horizon.
In view of this result, 10-dimensional black holes
deserve a more detailed study in which the $0$-brane description will
undoubtedly be a useful tool. 

In general, D-branes are an ideal
black hole laboratory because they naturally create heavy black holes
in the weak coupling limit, with masses of order $1/g$ 
(for comparison, in order to make a black hole
out of a long fundamental string, one needs to wrap it many times
around a non-contractible cycle \cite{cmp}). 
In order to study black holes in four dimensions we may toroidally 
compactify the ten-dimensional theory. A Dirichlet 6-brane wrapped
around the 6-torus is then a 4-dimensional black hole,
whose properties may be studied by scattering massless states off
of it.  We postpone a 
detailed discussion of these fascinating issues for future work.

\section*{Acknowledgements}

We are grateful to C.~Callan and L.~Thorlacius for illuminating
discussions.  This work was supported in part by DOE grant
DE-FG02-91ER40671, the NSF Presidential Young Investigator Award
PHY-9157482, and the James S.  McDonnell Foundation grant No.  91-48.
S.~Gubser was supported by the Hertz Foundation.

\end{document}